\documentclass{aastex61}

\submitjournal{PASP}

\shorttitle{RAVS}
\shortauthors{Xu et al.}

\begin{document}

\title{A Real-time Automatic Validation System for Optical Transients detected by GWAC}

\correspondingauthor{L. P. Xin}
\email{xlp@bao.ac.cn}

\author[0000-0002-0786-7307]{Y. Xu}
\affil{CAS Key Laboratory of Space Astronomy and Technology, National Astronomical Observatories, Chinese Academy of Sciences, Beijing, 100012, China}
\affiliation{University of Chinese Academy of Sciences, Beijing, 100049, China}

\author{L. P. Xin}
\affiliation{CAS Key Laboratory of Space Astronomy and Technology, National Astronomical Observatories, Chinese Academy of Sciences, Beijing, 100012, China}

\author{J. Wang}
\affiliation{Guangxi Key Laboratory for Relativistic Astrophysics, School of Physical Science and Technology, Guangxi University, Nanning 530004, Peoples Republic of China}

\author{X. H. Han}
\affiliation{CAS Key Laboratory of Space Astronomy and Technology, National Astronomical Observatories, Chinese Academy of Sciences, Beijing, 100012, China}

\author{Y. L. Qiu}
\affiliation{CAS Key Laboratory of Space Astronomy and Technology, National Astronomical Observatories, Chinese Academy of Sciences, Beijing, 100012, China}

\author{M. H. Huang}
\affiliation{CAS Key Laboratory of Space Astronomy and Technology, National Astronomical Observatories, Chinese Academy of Sciences, Beijing, 100012, China}
\affiliation{University of Chinese Academy of Sciences, Beijing, 100049, China}

\author{J. Y. Wei}
\affiliation{CAS Key Laboratory of Space Astronomy and Technology, National Astronomical Observatories, Chinese Academy of Sciences, Beijing, 100012, China}
\affiliation{University of Chinese Academy of Sciences, Beijing, 100049, China}



\begin{abstract}

The ground-based wide-angle camera array (GWAC) generates millions of single frame alerts per night. After the complicated and elaborate filters by multiple methods, a couple of dozens of candidates are still needed to be confirmed by follow-up observations in real-time. In order to free scientists from the complex and high-intensity follow-up tasks, we developed a Real-time Automatic transient Validation System (RAVS), and introduce here its system architecture, data processing flow, database schema,  automatic follow-up control flow, and mobile message notification solution.  
This system is capable of automatically carrying out all operations in real-time without human intervention, including the validation of transient candidates, the adaptive light-curve sampling for identified targets in multi-band, and the pushing of observation results to the mobile client. The running of RAVS shows that an M-type stellar flare event can be well sampled by RAVS without a significant loss of the details, while the observing time is only less than one-third of the time coverage. 
Because the control logic of RAVS is designed to be independent of the telescope hardware, RAVS can be conveniently transplanted to other telescopes, especially the follow-up system of SVOM. Some future improvements are presented for the adaptive light-curve sampling, after taking into account both the brightness of sources and the evolution trends of the corresponding light-curves.

\end{abstract}

\keywords{methods: observational --- telescopes --- surveys}



\section{Introduction} \label{sec:intro}

The ground-based wide-angle camera array (GWAC) is a set of ground-based instruments under the framework of the SVOM mission \citep{2015arXiv151203323C,2016arXiv161006892W}. SVOM, funded by the China National Space Administration and the Centre National d’Etudes Spatiales of France, is a Chinese–French space mission dedicated to detecting gamma-ray bursts (GRBs) and is planned to launch in 2021. The scientific motivation of GWAC is to detect the optical prompt emission of GRBs by simultaneously monitoring an area of sky within the field-of-view (FoV) of ECLAIRs that is the major gamma-ray trigger instrument of SVOM. With this motivation, GWAC is designed to comprise 36 cameras, each with an 18 cm diameter and $12.4 \times 12.4\ deg^2$ FoV. The 36 cameras point to the sky in different directions and totally cover an area of more than 5000 $deg^2$ at the same time. Each camera takes an image once every 15s (10s exposure plus 5s readout) with a limit magnitude of $m_r = 16 mag$ in a moonless night. Thanks to its large FoV and short survey cadence, besides catching GRBs, GWAC is also an optical transient factory in discovering diverse optical transients, including supernovae, tidal disruption events, dwarf novae, stellar flares and the optical counterparts of both gravitational-wave bursts and neutrino events. For example, three dwarf novaes \citep{2020AJ....159...35W} are identified independently by GWAC. Two identical 60 cm diameter optical telescopes (GWAC-F60A/B \footnote{GWAC-F60A/B was co-founded by Guangxi University and NAOC. The limit magnitude is 17.5mag in R band @30sec.  The mount is bought from ASA company.  The standard Johnson-basal  filters, U,B,V,R,I and clear filters are used.}, short for F60s) and a 30 cm diameter optical telescope (GWAC-F30 \footnote{GWAC-F30 was co-founded by Huaibei Normal University and NAOC. The limit magnitude is 16.0mag in R band @30sec. The mount is also bought from ASA company.  The filters are the same as GWAC-F60A/B.}) that are located at the dome of GWAC are equipped for follow-up observations. We refer the readers to \citep{2020RAA....20...13T} for a more detailed description of GWAC.

Similar to Zwicky Transient Facility (ZTF) \citep{2020PASP..132c8001D} and other wide-field survey projects, one key issue in GWAC is how to pick up real transients from millions of alerts in one night by auto filtering and by subsequent follow-up observations. The auto filtering has been implemented by a dedicated pipeline including a series of filters based on morphological parameters, and machine learning algorithms (Xu et al. in preparing), and so on. After the filtering, there are typically still a couple of dozens of candidates that are necessary to be confirmed by follow-ups, which is infeasible for manual observations and identifications because of the following reasons. At first, we do not know when an alert is generated in advance. Secondly, an optimal follow-up strategy is quite hard to be promptly made by a duty scientist by hand when there are multiple active events, each with a time scale from minutes to hours. 

In fact, contemporaneous observations by different types of telescopes and instruments at a range of wavelengths are essential to understand diverse important astrophysical phenomena, e.g., young supernovae \citep{2015Natur.521..328C}, gamma-ray bursts \citep{2015ApJ...803L..24C}, supernovae \citep{2014Natur.509..471G}, etc. In order to shorten the delay of follow-ups and to follow-up multiple events by an optimal strategy, an auto follow-up system must be established to avoid overloading of a duty scientist and to guarantee the scientific return. The system is able to not only distribute a filtered alert to follow-up telescopes, which is similar to the function of the ZTF Alert Distribution System \citep{2019PASP..131a8001P}, but also make an optimal follow-up strategy and inform the follow-up result to duty scientist by multiple ways in real-time. 
The traditionally used ways of message delivery are e-mail and PC based web service, which is hard for real-time communication. With the rapid development of mobile chipset and wireless network technology, mobile phone evolves from a simple communication terminal to a mobile computer with powerful functions which is the best convenient way of real-time message acquirement.

In this paper, we describe the Real-time Automatic transient Validation System (RAVS) of GWAC. This system can implement all the three functions mentioned above, i.e., alert distribution, strategy generation, and result notification. The organization of the paper is as follows. In Section \ref{sec:system}, we briefly introduce the overview of this system, including the system architecture, data stream, and schema of the database. Section \ref{sec:strategy} describes the control pipeline of RAVS in details, and shows the automatic validation process of a fast transient as an example. Section \ref{sec:mobile} describes the mobile notification solution of RAVS. In Section \ref{sec:future}, we discuss future research directions of RAVS. Finally, we make a summary in Section \ref{sec:summary}.

\section{System Overview} \label{sec:system}

To free astronomers from the complex follow-up observation job, reduce the follow-up delay and improve the observation efficiency of F60s, we designed and realized the Real-time Automatic transient Validation System (RAVS) of GWAC. RAVS is a fully automatic follow-up observation control system that automatically confirms transient candidates detected by GWAC by rapid follow-up observation, performs multi-band photometry for the confirmed transient and pushes observation results to duty scientists in real-time. And all of these operations are performed without human intervention.

\subsection{System Architecture and Data Processing Stream} \label{subsec:sysArch}

\begin{figure}
\plotone{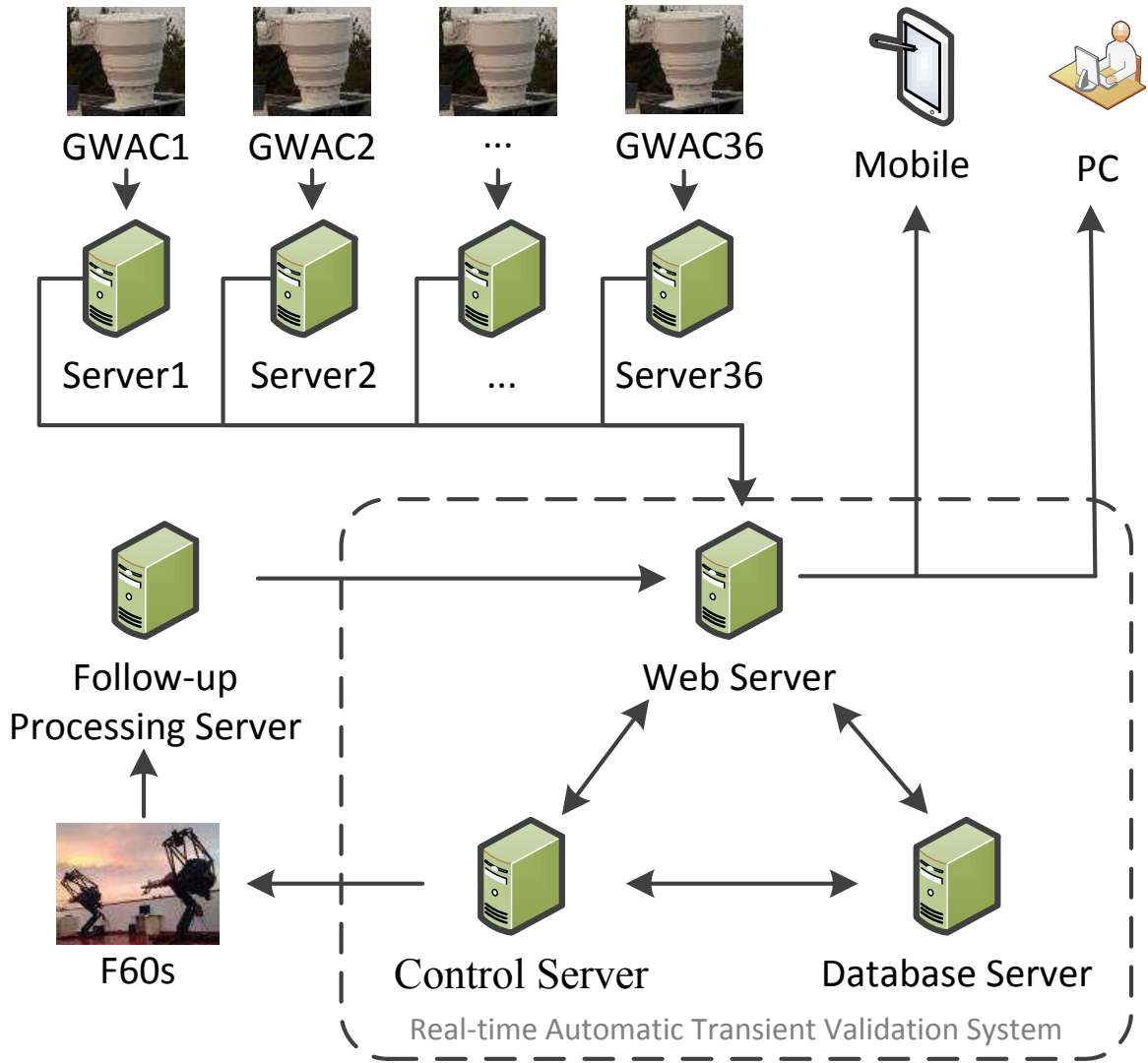}
\caption{System Architecture and data processing stream. The RAVS is in the dashed-line box. The data flow is shown by the arrows.\label{fig:sysArch}}
\end{figure}

Figure \ref{fig:sysArch} shows the system architecture and data flow of RAVS. It is composed of three servers: the control server, the web server and the database server. The latter two servers are shared with other subsystems of GWAC. The RAVS automatically completes the subsequent follow-up tasks through the following data flow:

\begin{enumerate}
\item At the beginning, the web server receives a transient candidate generated by the GWAC cameras, transfers the corresponding data to the database server, and sends a message to the control server by a message queue;
\item After receiving the message, the control server queries the database to obtain the detailed information of the candidate. Meanwhile, a follow-up request is generated and sent to F60s by the control server, and the related information of the candidate is updated in the database; 
\item Once the follow-up observation and the following data process are finished, the photometry results are received and then transferred to the database by the web server;
\item Control server queries the photometry results in the database every 5 seconds and automatically carries out the next observation according to the latest photometry results;
\item For a confirmed transient, the alarm message is distributed to the PC and mobile phone through the Web server.
\end{enumerate}

\subsection{Database Design} \label{subsec:database}

\begin{figure}
\plotone{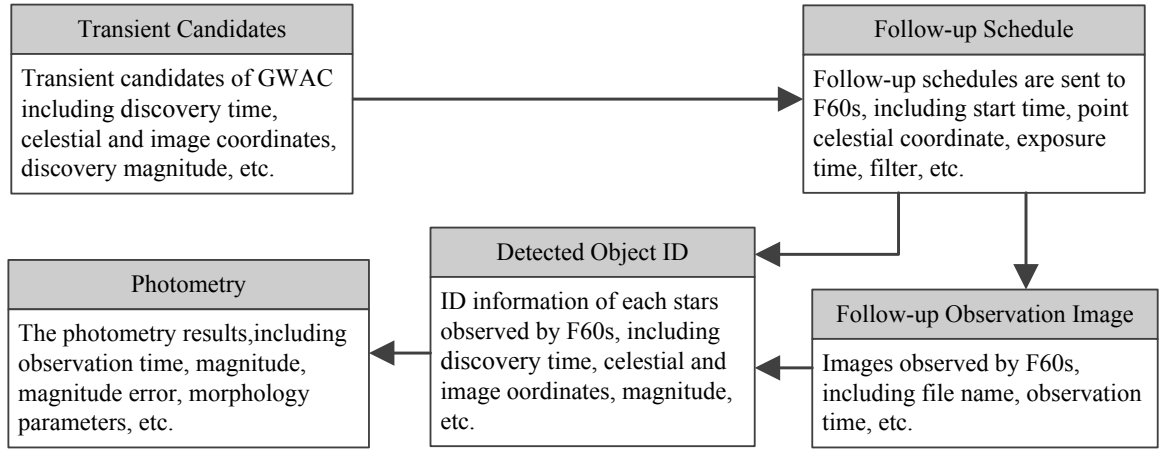}
\caption{Simplified key schema of RAVS database.\label{fig:dbDesi}}
\end{figure}

In order to support the data processing of RAVS, we create five tables: transient candidates table, follow-up schedule table, follow-up observation image table, detected object ID table and detected object record table. These five tables are briefly described as follows and the related structures are shown in Figure \ref{fig:dbDesi}.

\begin{enumerate}
\item Transient candidates table: It stores the transient candidates generated by the GWAC auto filtering;
\item Follow-up schedule table: This table stores the follow-up schedule of each received candidate. The schedule for each candidate is adaptive by dynamically adjusting the exposure time, filter and cadence based on the evolution of its multi-band light curves. More detailed information will be presented in Section 3; 
\item Follow-up observation image table: This table stores the information of the observed images including observation time, image name and store path, etc;
\item Detected object ID table: Each detected new object is assigned a unique ID;
\item Photometry table: This table stores all the photometry results of those detected objects, such as observation time, magnitude, magnitude error, filter, coordinates, morphology parameters, etc.
\end{enumerate}

\section{Adaptive Strategy of RAVS} \label{sec:strategy}

\begin{figure}
\plotone{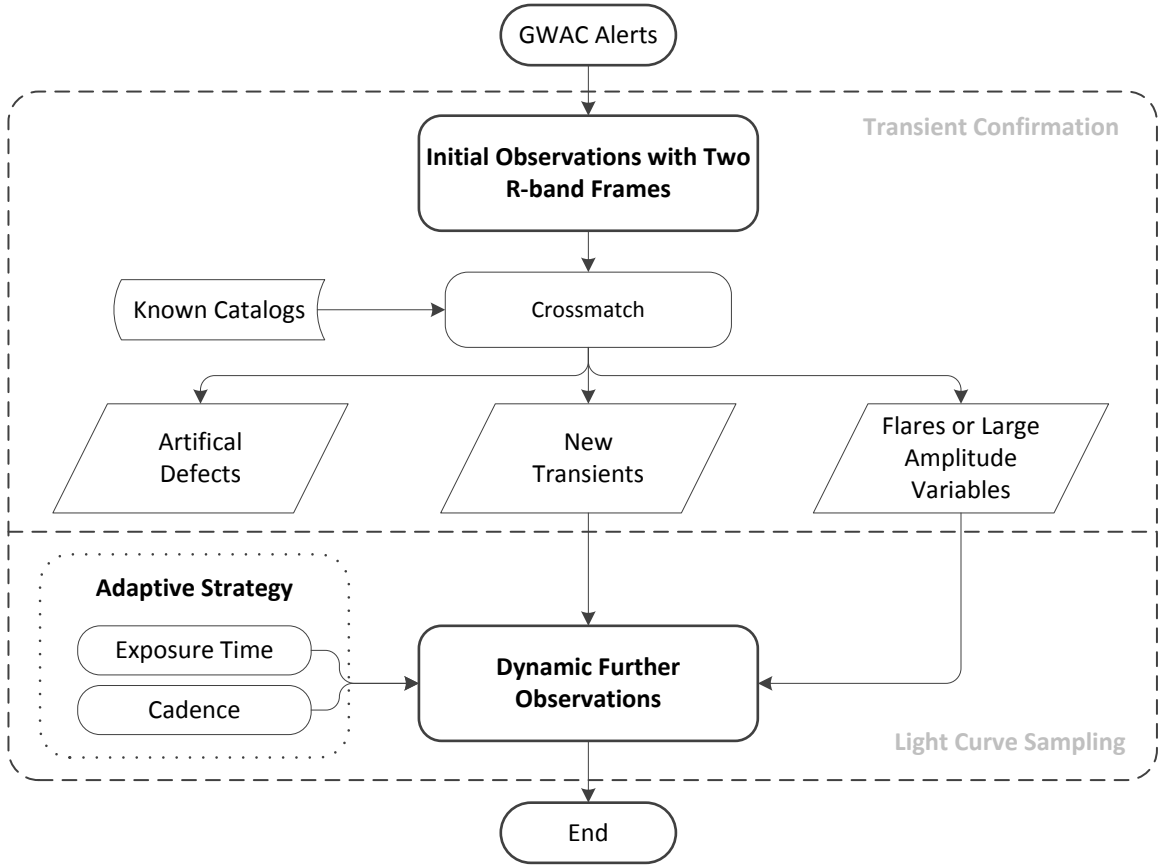}
\caption{RAVS’s control pipeline including the confirmation stage and the light curve sampling stage.\label{fig:strategy}}
\end{figure}

The automatic flow chart of RAVS is briefly shown in Figure \ref{fig:strategy}. The follow-up observation of a transient candidate can be divided into two steps. The first is candidate confirmation, and the second light curve sampling for the confirmed transients. In order to improve the efficiency of follow-up observation, it is necessary to dynamically adjust the exposure time, filter and cadence based on the evolution of multi-band light curves. The detailed strategies in both stages are described in the following sections of \ref{sec:transConf} and \ref{sec:curveSamp}.

\subsection{The confirmation of transient candidate} \label{sec:transConf}

The main purpose of the confirmation stage is to confirm candidates by F60s that have much higher spatial resolution and deeper sensitivity compared to GWAC. F60s can be automatically triggered for each candidate found by GWAC with a short delay. Each candidate is observed in R band by F60s twice successively, in order to save the rare cases that the first image could not be usable because of the instability of the telescope after a very fast slew. For each frame, the exposure time for F60s increases to 30 seconds from 10 sec for GWAC. The enhanced detection limit by about 3 magnitudes ensures that a rapidly fading candidate is still detectable by F60s after one minute. 

A real-time data processing pipeline dedicated to the confirmation has been developed based on cross-matching of the candidates with catalogs, such as USNO B1.0 \citep{2003AJ....125..984M} , 2MASS \citep{2003tmc..book.....C} , Gaia Dr2 \citep{2016A&A...595A...1G,2018A&A...616A...1G}. With the pipeline, the confirmation results in the following three types of objects: 

\begin{enumerate}
\item Artificial defects: F60s do not detect any objects at 5 sigma significance level within 30 arcseconds around the candidate positions determined by GWAC. In this case, most of them are hot spots of GWAC detectors, ghosts, moving targets, etc. However, the possibility of targets with very rapidly fading cannot be completely ruled out.
\item Flares or large-amplitude variables: A candidate is classified as either a flare star or a large-amplitude variable in the following conditions. At first, F60s detect at least one source within the same area. Secondly, each source has a counterpart within 3 arcsec in known catalogs, such as USNO B1.0, 2mass, Gaia Dr2. Finally one of them is brighter by more than one magnitude compared to the catalogs. 
\item New transient: Some of source detected by F60s within 30 arcseconds around the candidate positions determined by GWAC have no counterpart within 3 arcsec in the known catalogs.
\end{enumerate}

\subsection{The light curve sampling} \label{sec:curveSamp}

Once a transient is confirmed, an automatic and flexible exposure sequence of light curve sampling taken by F60s is necessary for obtaining maximum scientific returns with minimum costs after taking into account of the following two requirements.

\begin{enumerate}
\item A balance  of multi-tasks. There are several active exposure sequences for F60s in parallel. These tasks include confirmations of other candidates, observations of GRB afterglows and supernovae. 
\item A balance  of exposure time and cadence. In principle, the exposure time depends on the brightness, and the cadence depends on the evolution of the light curve. A long exposure results in a low cadence, while a high cadence needs a short exposure. 
\end{enumerate}

Based on these considerations, we dynamically adjust the exposure time and the delay time in the pipeline of RAVS to obtain a relatively complete light curve with as few costs as possible. The adjustment is mainly based on a prediction of the brightness according to an extrapolation of the observed light curve. For a given exposure sequence ($T_{i+1},t_{i+1},m_{i+1}$ where $i=1,2,3,…,n-1,n…$), the exposure time $t_{i+1}$ and the observations time $T_{i+1}$ can be determined from the finished sequence ($T_i,t_i, m_i$ where $i=1,2,3,…,n-1,n…$), where $m_i$ is the magnitude in the $i$th frame. The details are described in the next two subsections.

\subsubsection{Exposure time}

For a target, the formula for calculating the exposure time of the $(n+1)$th observation is:
\begin{eqnarray}
t_{n+1} & = & t_n \times 10^{0.4 \Delta m_n}, n\geq3, t_n\geq3\ seconds
\end{eqnarray}
where $\Delta m_n=m_n-m_{n-1}$. The maximum exposure time of 200 seconds is setup considering the tracking stability of F60s.

\subsubsection{Observation time}

The $(n+1)$th observation time $T_{n+1}$ is determined from
\begin{eqnarray}
T_{n+1} & = & T_n + \delta T_{n+1}
\end{eqnarray}
where $T_n$ is the $n$th observation time. $\delta T_{n+1}$is the delay time between two exposures which is determined by the following formula: 
\begin{eqnarray}
\delta T_{n+1} & = & \left\{
\begin{array}{lr}
0, & \Delta m_f\geq0.3 \\
3\ \mathrm{minutes}, & \Delta m_f<0.3, \Delta m_s\geq0.1 \\
(1+f)(T_n-T_m), & n\geq 3, m<n, \Delta m_f<0.3, \Delta m_s<0.1, f=0.3
\end{array}\right.
\end{eqnarray}
where $1+f$ is a scale factor of delay time,  $\Delta m_f=|m_n-m_{n-1}|$,  $\Delta m_s=|m_n-m_m|$, $m_m$ is the magnitude at the $m$th observation time $T_m$. The $T_m$ is the start time at which the light curve evolves into a flat phase.  The index m is determined by the formula:
\begin{eqnarray}
m & = max(i), |m_n-m_{n-1}|<0.3, |m_n-m_{i-1}|>0.1, i\geq2
\end{eqnarray}

\begin{figure}
\centerline{
\includegraphics[width=15cm]{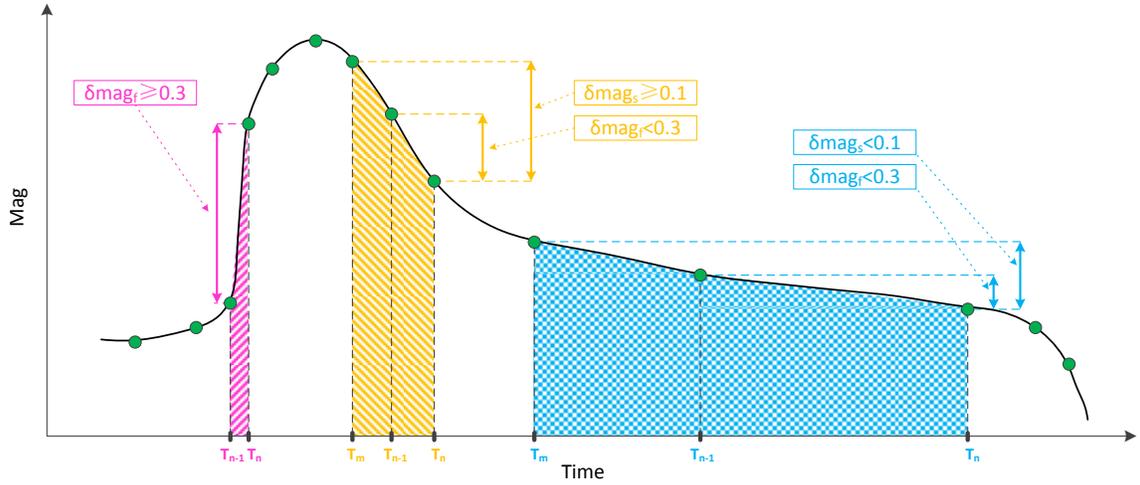}}
\caption{
The above figure is a typical schematic diagram of the evolution of light curve which is divided into three kinds of trends corresponding to the equation (3). First, the pink shadow on the left of the figure represents the stage changing very quickly, in which the observation delay is zero. Second, the yellow shadow in the middle represents the quick change stage having an observation delay of three minutes. At last, the blue shadow on the right represents the stage changing relatively slow, in which the observation delay is dynamically adjusted.
\label{fig:lightCurve1}}
\end{figure}

The maximum values of $\delta T_n$ and $T_n$ are 40 minutes and 180 minutes respectively. The adaptive exposure strategy described above is schematically illustrated by a cartoon in Figure \ref{fig:lightCurve1}.

\subsection{Example}

\begin{figure}
\plotone{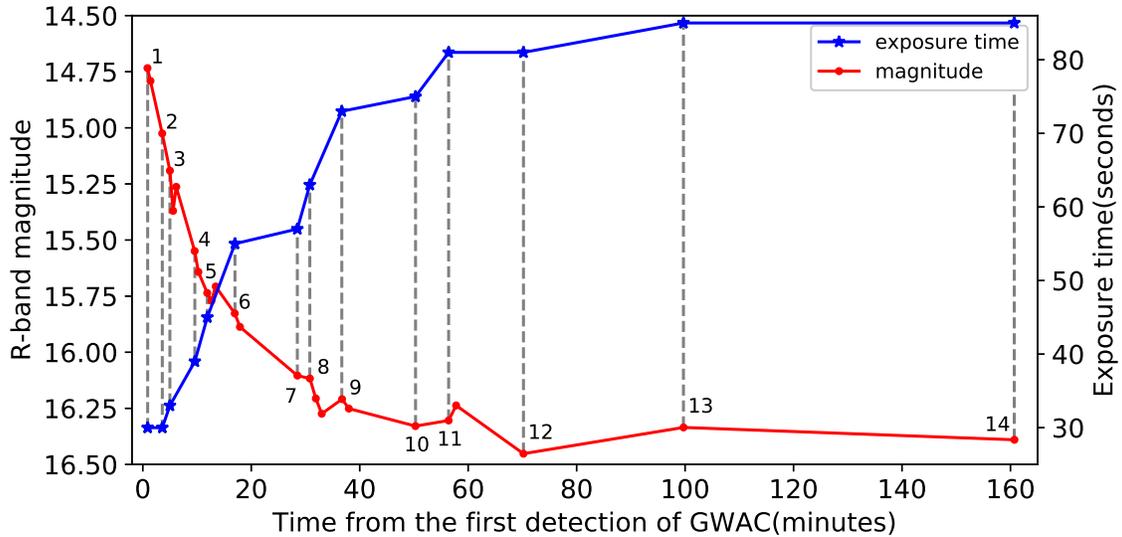}
\caption{Observation detail of fast transient GWAC\,190101A. Red line with point mark is the light curve observed automatically by RAVS. Blue line with star mark is the exposure time line which is dynamically calculated by equation (1) along with the change of light curve. Each gray dashed line with one endpoint connecting point mark of red line and another endpoint connecting star mark of blue line represent one observation automatically triggered by RAVS, and each is labeled with a number which is the observation sequence. The horizontal space between gray dashed is the exposure delay which is dynamically adjusted by equation (2). \label{fig:G190101C17639}}
\end{figure}

Figure \ref{fig:G190101C17639} shows an example of the follow-up results of fast transient GWAC\,190101A sampled by RAVS, which is finally identified as a large amplitude flare of an M-dwarf. The red and blue symbols denote light curve and exposure time curve, respectively. Additionally, two facts can be learned from the diagram. On the one hand, faster the variation, higher the cadence will be. On the other hand, the exposure time increases with the magnitude.

For this example, the total light-curve covers a duration of 160 minutes, in which the number of the observation request is 14. All the time consumed by these observations including the slew and exposure time of F60 is about 43 minutes which is about 27\% of the full light-curve time coverage. One can learn from this example that RAVS could save 73\% of the observation time of F60 without significant loss of the details of the light-curve.

\section{Mobile Notification} \label{sec:mobile}

At present, there are a variety of message delivery software worldwide, such as Slack, WeChat, Skype, WhatsApp, and Telegram, etc. RAVS uses enterprise WeChat\footnote{https://work.weixin.qq.com} as a solution of the real-time mobile notification for pushing follow-up  results of each alert and ancillary data. 
   
WeChat developed by Tencent company is the most popular message pushing software in China. WeChat has a billion users and is operated in Android, IOS, and other mobile operating systems. Enterprise WeChat provides a wealth of user and message management functions to support the transmission of text, images, files, voice, video and other types of information. Enterprise WeChat additionally provides a general development interface based on Web services, as well as management pages and fully functional mobile phone clients. Using enterprise WeChat, developers can enjoy stable and reliable services and only needs to focus on the processing and analysis of astronomical data, regardless of the differences of different mobile operating systems and the re-development requirements caused by the upgrade of mobile phone. The enterprise WeChat can provide multi-channel for message delivery, which enables us to push different types of messages on different channels. Each user of RAVS can subscribe channels for interesting messages. 

\begin{figure}
\plotone{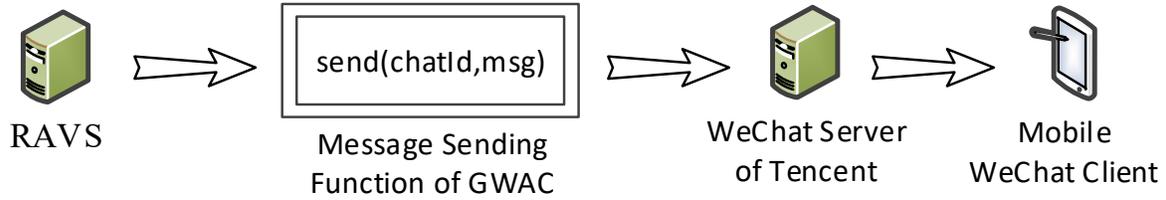}
\caption{Sample pipeline of sending message to Enterprise WeChat client. All the processes of calling WeChat interface are encapsulate as a function:\texttt{send(chatId,msg)}. \label{fig:mobile}}
\end{figure}

Figure \ref{fig:mobile} shows the process of sending messages to the mobile terminal via enterprise WeChat. RAVS encapsulates the message sending interface and the relative parameters  as a python function: \texttt{send(chatId,msg)}, where \texttt{chatId} is channel ID and  \texttt{msg} is the message in text or figure format.

\begin{table}[h!]
\renewcommand{\thetable}{\arabic{table}}
\centering
\caption{Messages of follow-up result reported to astronomer} \label{tab:mobileMsg}
\begin{tabular}{lcl}
\tablewidth{0pt}
\hline
\hline 		
Item & Type & Contents \\
\hline
\decimals
NAME & String & Name of the confirmed candidate  \\
TYPE & String & Candidate type  \\
STAGE & Integer & Stage of the full followup pipeline  \\
USNO R2 & Float & R2 magnitude of USNO B1.0 catalog  \\
USNO B-R & Float & B2-R2 color of USNO B1.0 catalog  \\
MAG\_GWAC & Float & Detection magnitude by GWAC in white band  \\
MAG\_60\_FIRST & Float & The magnitude of the first observation by F60s in R-band  \\
MAG\_60\_LATEST & Float & The magnitude of the latest observation by F60s in R-band  \\
MAG\_DIFF\_GLOBAL & Float & mag\_60\_latest - mag\_60\_first  \\
MAG\_DIFF\_LOOP & Float & The difference of magnitude during the current stage  \\
DETAIL\_URL & String & The URL of GWAC web page  \\

\hline
\end{tabular}
\end{table}

\begin{figure}
\centerline{
\includegraphics[height=8cm]{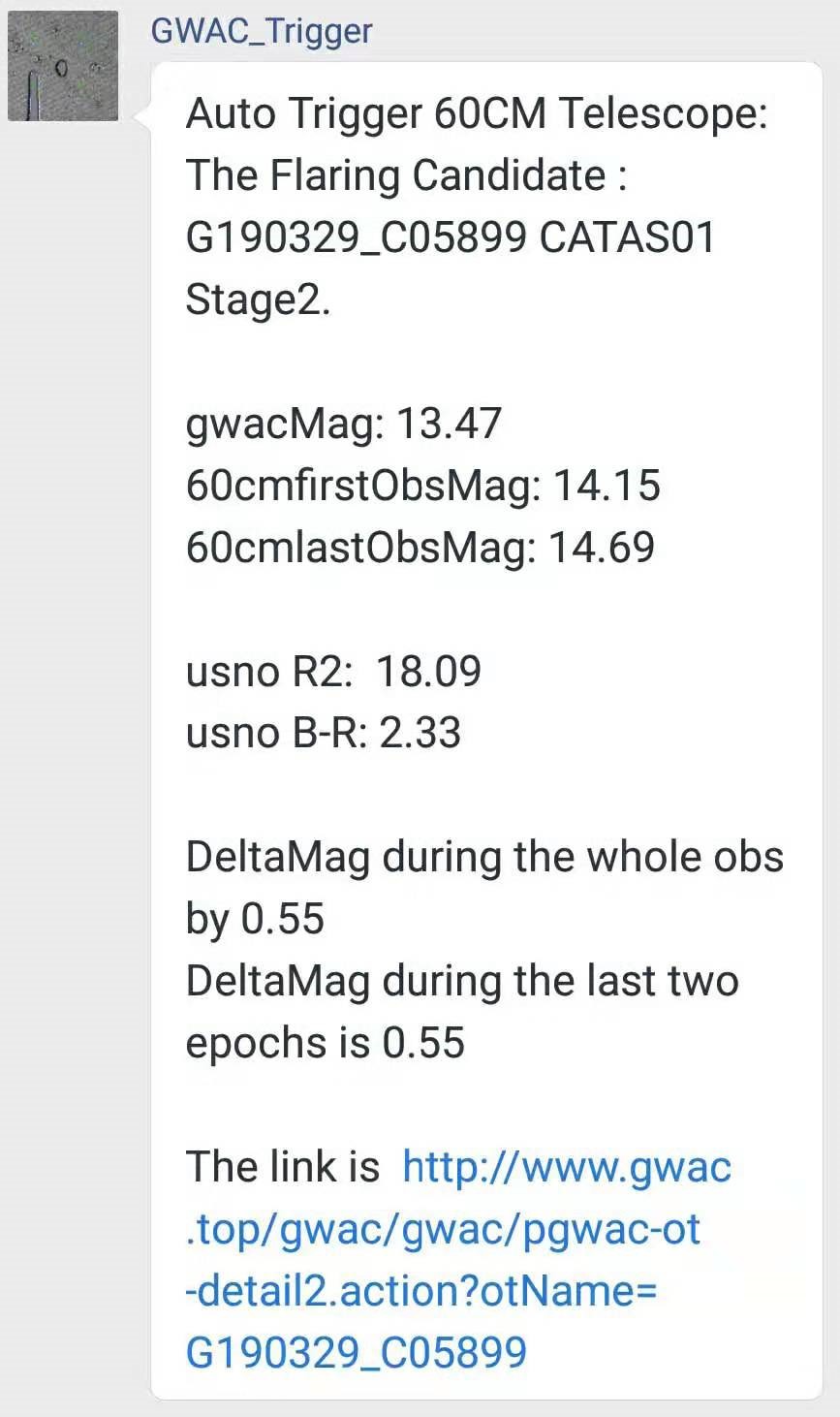}
\includegraphics[height=8cm]{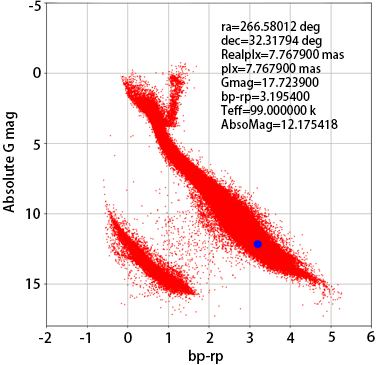}
}
\caption{Real-time WeChat notification about the digest information (left) and Hertzsprung Russell diagram (right) of a flare of GWAC.\label{fig:mobileInfo}}
\end{figure}

As an example, Table \ref{tab:mobileMsg} lists the entries pushed to \texttt{All Transient} channels. Figure \ref{fig:mobileInfo} is a snapshot of Wechat for one transient identified as a flare of an M-dwarf. 
The left panel shows the summary of one exposure. The blue point in the right panel of the figure marks the corresponding position in the Hertzsprung-Russell diagram that is built from the Gaia Dr2 catalog. 

\section{Project status and future work} \label{sec:future}

The control pipeline of RAVS is written in Python and uses Postgresql as the archive database. The reduction software for the processing of follow-up images is developed by calling SExtractor \citep{1996A&AS..117..393B}, iraf \citep{1986SPIE..627..733T}, etc. 

Currently, there are 8 telescopes of GWAC in operation and others under maintenance. We made statistics on the data of January 2020, and a total of 14 nights were observed for more than 8 hours. The average number of single frame alerts of those nights is about 400,000. 0.03\% of those alerts are identified as transient candidates after the filters of morphological parameters, machine learning algorithms, etc. Candidates are chronologically followed by RAVS without priority difference, and preferentially re-followed with a higher priority if the confirmed targets change acutely, otherwise re-followed by chronological order with priority not changed. 12.8\% of those followed are automatically confirmed to be real transient that will be manually checked by duty scientists subsequently. To date, GWAC has detected about 150 flare events, including three published dwarf novaes \citep{2020AJ....159...35W}.

The followed and re-followed observation commands are sent to the global scheduler of GWAC system. This scheduler is separated from RAVS and designed for performing observations for all GWAC scientific objectives by scheduling all telescopes. The scheduling algorithm is mainly based on the priority of the target. Other astronomic parameters, such as zenith, hour angle, and so on, are also considered as secondary factors (Han et al. in preparing).

The RAVS telescopes (F60s) are carrying on surveys, queue observations of user applied objects, follow-ups of LIGO/Virgo GWs and Swift/Fermi GRBs. For each type of object, several specific observation strategies are adopted to deal with different scenarios and the priority is pre-defined for each case. For RAVS targets, the scheduler maintains an observation sequence in order of priorities and arriving time. For other types of targets, the scheduling strategies can be different. In principle, the targets from RAVS have the top priority for one of F60s telescopes. The observations of these targets will not be interrupted by other types of targets. Cloud or haze may reduce the number of transient candidates, which decreases the follow-up pressure of F60s. We will stop the observation in worse weather.

The current RAVS can basically meet our scientific requirements, while there are still some functions to be optimized, which are listed as follows:
\begin{enumerate}
\item The exposure time in the confirmation stage: Currently, the initial exposure time for the transient confirmation is constantly set as 30 seconds for all candidates, regardless of their brightness reported by GWAC data processing. Actually, the magnitude observed by GWAC has been stored in the database. Thus, the exposure time in the confirmation stage for one given candidate will be set taking into account its brightness in the next version. In other words, more bright, less exposure time.
\item The observation sequence in the stage of light curve sampling: The exposure time of the next observation in the light curve sampling is calculated by the change of magnitudes in the last two observations. However, the optimized exposure time of the next observation shall be estimated by the fitting of all the observation data in history.
\item The robustness of RAVS:
RAVS will be paralyzed if the number of candidates from GWAC is too much to be followed by F60s promptly. A more intelligent scheduling needs to be developed, based on the priority, brightness, morphology of these candidates.
\end{enumerate}

\section{Summary} \label{sec:summary}

The Real-time Automatic transient Validation System (RAVS) of GWAC is introduced, which includes system architecture, data processing flow, database structure, the adaptive strategy and the mobile notification solution. With RAVS, one transient candidate detected by GWAC can be automatically validated in real-time by adaptive follow-up strategy without human involvement. As an example, an M-type stellar flare event was well sampled by RAVS without a significant loss of the details, while the observing time is only less than one-third of the time coverage. 
In the statistics of January 2020, 12.8\% of the follow-up observed candidates is automatically confirmed to be real transient, and most are real flares checked by duty scientists subsequently.
RAVS can be conveniently ported to other telescopes, especially the follow-up system of SVOM, since the control logic of RAVS is designed to be independent on the telescope hardware. RAVS still has some space to be improved in the adaptive strategy by taking into account both the brightness observed by GWAC and the evolution trend of the corresponding light-curve.

\section{Acknowledgement}

I would like to thank the members of the GWAC team for their suggestions and help during the experiment. 
This work is supported by the National Natural Science Foundation of China (Grant No. 11533003, 11973063, U1531134).
This work has made use of data from the European Space Agency (ESA) mission Gaia (https://www.cosmos.esa.int/gaia), processed by the Gaia Data Processing and Analysis Consortium (DPAC, https://www.cosmos.esa.int/web/gaia/dpac/consortium). Funding for the DPAC has been provided by national institutions, in particular the institutions participating in the Gaia Multilateral Agreement.

\end{document}